# Local probing of propagating acoustic waves in a gigahertz echo chamber

Martin V. Gustafsson[†], Paulo V. Santos[††], Göran Johansson[†] and Per Delsing[†]

[†]Microtechnology and Nanoscience, Chalmers University of Technology, Göteborg, Sweden

[††]Paul-Drude-Institut für Festkörperelektronik, Berlin, Germany

**In the same way that micro-mechanical resonators resemble guitar strings and drums, Surface Acoustic Waves (SAW) resemble the sound these instruments produce, but moving over a solid surface rather than through air. In contrast with oscillations in suspended resonators, such propagating mechanical waves have not before been studied near the quantum mechanical limits. Here, we demonstrate local probing of SAW with a displacement sensitivity of 30am$_{RMS}$/√Hz and detection sensitivity on the single-phonon level after averaging, at a frequency of 932MHz. Our probe is a piezoelectrically coupled Single Electron Transistor, which is sufficiently fast, non-destructive and localized to let us track pulses echoing back and forth in a long acoustic cavity, self-interfering and ringing the cavity up and down. We project that strong coupling to quantum circuits will allow new experiments, and hybrids utilizing the unique features of SAW. Prospects include quantum investigations of phonon-phonon interactions, and acoustic coupling to superconducting qubits, for which we present favourable estimates.**

Surface acoustic waves exist on scales ranging from the microscopic to the seismic and resemble ripples on a pond, but traversing the surfaces of solids. SAW with wavelengths down to the sub-micrometer range can be efficiently generated and controlled on a piezoelectric chip, using lithographically fabricated transducers and gratings. The waves experience low losses during propagation and reflection, allowing them to be harnessed for a variety of electro-mechanical microwave applications, including acoustic delay lines, resonators and filters[1,2]. Conventional photolithography limits the frequency of commercial SAW devices to around $f$=3GHz, but much higher frequencies can be obtained using more advanced techniques[3].

When one aims to control individual quanta, a high operating frequency $f$ is essential regardless of the specific system, since the quantum energy $hf$ (where $h$ is Planck's constant) must well exceed the thermal energy ($hf>>k_BT\approx$20mK$\times k_B\approx$400MHz$\times h$ for conventional cooling methods). In recent pioneering experiments, the displacements of string-like nanomechanical resonators have been measured near the quantum mechanical limit[4,5] and drum-like resonators have been made to vibrate quantum-coherently with a superconducting qubit[6] and couple strongly to an electrical resonator[7].

To reach sufficiently high displacement sensitivity to detect single phonons in a mechanical system, it is desirable to keep the vibration sustained, *i.e.* use a resonator with high quality factor $Q$. For frequencies in the range of several hundreds of MHz, SAW devices can have quality factors above $10^5$ when operated at low temperature[8,9], whereas suspended micro-resonators typically do not exceed $10^3$ [10] with recently demonstrated carbon-based devices as notable exceptions[11,12]. This indicates that SAW technology has useful potential in the field of quantum phononics merely through its ability to confine high frequency phonons. As we will show, however, detection sensitivity on the single-phonon level is possible also without strong cavity confinement, *i.e.* for freely propagating mechanical states.

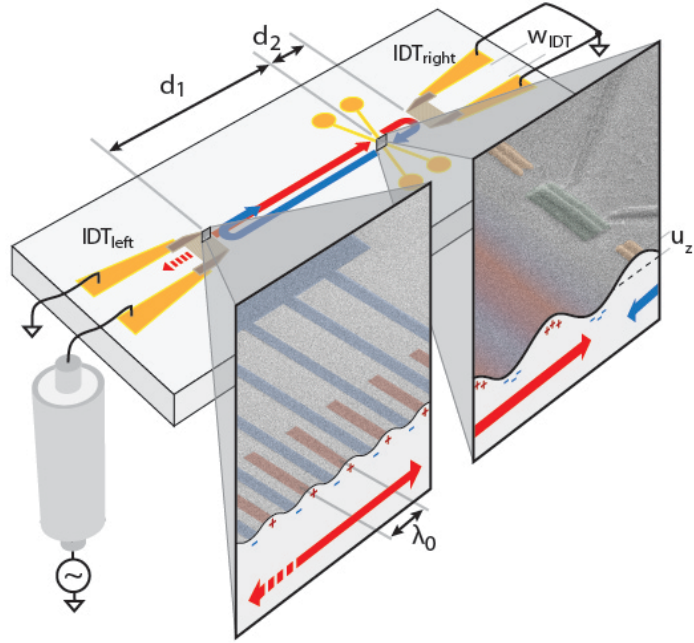

**Figure 1: Sample layout**

**Overview of the sample layout, with zoom-in SEM micrographs in false-color of an Interdigital Transducer (IDT) and the SET. The SET is deposited in-between the two IDTs, at a distance $d_1$=2252µm from IDT$_{left}$ and $d_2$=651µm from IDT$_{right}$. The IDTs are lithographically identical, each with 160 fingers of length $W_{IDT}$=320µm and a period of $\lambda_0$=3.12µm. IDT$_{left}$ is connected to an RF source, and generates SAW equally in the left and right direction (left micrograph) when excited by a signal near the IDT resonance frequency. When the right-going wave (solid red arrow) meets the grounded IDT$_{right}$, it reflects back towards IDT$_{left}$ (blue arrow), where it reflects again. In this way, an acoustic cavity mode builds up between the two IDTs. The right micrograph shows the SET, with the island colored green and the gate leads yellow. The island has an area of $A_{SET}$=1.4×0.4µm$^2$, and when the SAW passes underneath it, the island is polarized by the piezoelectric charge in this area. The polarization is proportional to the vertical displacement $u_z$ (magnified by ~10$^{10}$ for the purpose of illustration).**

We detect the SAW by coupling the piezoelectric charge it produces directly (*i.e.* without intermediate circuit elements) onto the island of a Single Electron Transistor (SET)[13]. The SET is the most sensitive electrometer in existence[14], and has been used to detect ultra-small displacements through modulation of a capacitance[15-17]. In our device, the SET island is deposited directly on a GaAs substrate, picking up the full polarization charge $q_{SAW}=A_{SET}e_{14}s_{xx}$, where $A_{SET}$ is the area of the SET island, $e_{14}$ is the substrate's piezoelectric constant, and $s_{xx}$ the surface strain caused by the acoustic wave along its direction of propagation.

Knobel and Cleland[18] proposed a similar scheme as a read-out for suspended-beam resonators already in 2002, but no experiments have been reported to date, even though this technique was predicted to yield much higher displacement sensitivity than capacitance modulation[19]. SAW devices are particularly well suited for this type of direct charge coupling, which was first utilized in attempts to realize a quantized current source[20], since the motion takes place at a flat and stiff surface where an SET or a similar device can easily be deposited.

The sample used in this study is illustrated in Fig. 1. It was fabricated on the [001] surface of a GaAs chip, where two aluminium Interdigital Transducers (IDTs) are deposited, facing each other. The IDTs are aligned with the [110] direction of the crystal, which has a SAW velocity of $v_{SAW} \approx 2900$m/s at low temperature. They are separated by 2903μm, and both have a finger period of $\lambda_0 = 3.12$μm. The finger length of $w_{IDT} = 320$μm sets the SAW beam width. The SET was deposited in the center of the beam, asymmetrically between the two IDTs.

$IDT_{left}$, which was connected to a microwave source with pulsing capability, produces a monochromatic SAW beam when driven at a frequency $f_{SAW}$ near the IDT resonance $f_0 = v_{SAW}/\lambda_0$. From electrical reflection measurements, we find that only 40% of the applied power is dissipated in the IDT on resonance. Assuming that this power is fully converted into SAW and emitted equally in the left and right directions, no more than a fraction $\beta_{gen} = 20\%$ of the applied electrical power is launched towards the SET and $IDT_{right}$. $IDT_{right}$ is grounded in order to act as a SAW reflector.

For improved bandwidth and sensitivity, the SET was operated in the RF mode[21], *i.e.* embedded in an LC-circuit and probed with an RF signal. The reflection coefficient of the circuit is modulated by the induced charge $q_{ind}$ on the SET island, giving a transfer function $\Gamma_0(q_{ind})$ which is periodic in the electron charge $e$ (Fig. 2b, black). Throughout this Article, we denote the SET response by $\Gamma$, which is the absolute RF-SET reflection coefficient, normalized so that $\Gamma_0(q_{ind})$ fits in a [0,1] interval.

Charge can be induced on the SET island both by the SAW and by an external voltage $V_g$ applied to a gate capacitor $C_g$; $q_{ind} = q_g + q_{SAW}$ with $q_g = V_g C_g$. Since the SAW frequency is lower than the intrinsic bandwidth of the SET[22] ($BW_{SET} \approx 14$GHz), but much higher than the RF-SET's bandwidth ($BW_{LC} \approx 8$MHz), the SAW manifests as a smearing of the transfer function $\Gamma_0(q_{ind})$; see Fig. 2b. Once $\Gamma_0(q_{ind})$ is known, the corresponding transfer function $\Gamma(q_g, q_{SAW})$ in the presence of SAW can be calculated for any $q_{SAW}$ (See Supplementary Methods II). The same type of diagram is obtained also for an RF signal that couples capacitively to the SET with no mechanical intermediary, but time-resolved measurements clearly show that the coupling is acoustic, and that the direct electrical interaction between the IDT and the SET is negligible (see below). The agreement between the measured and calculated $\Gamma(q_g, q_{SAW})$ is very good, as shown in Fig. 2b and 2c. This gives us a calibration of the charge amplitude induced by the SAW, with respect to the voltage amplitude applied to the IDT.

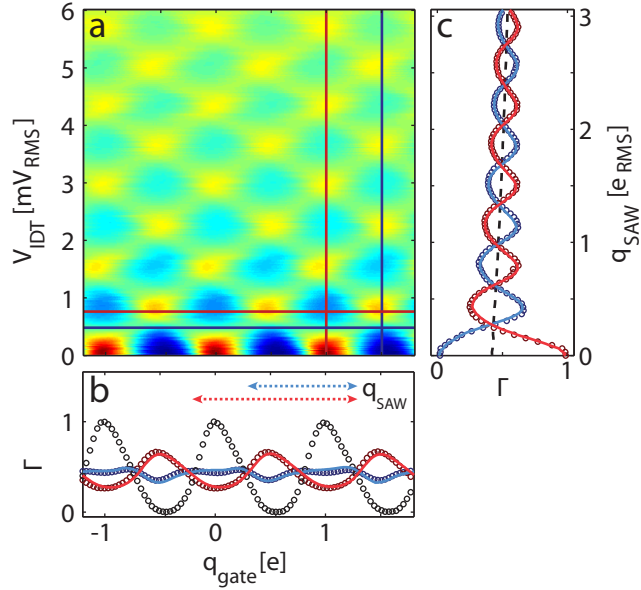

**Figure 2: Response of the SET to SAW in the steady state**

All three panels show the normalized RF-SET reflection coefficient $\Gamma$. The data were collected for a continuous voltage amplitude $V_{IDT}$ applied to $IDT_{left}$ at the frequency $f_{SAW}$ = 932.407MHz (blue dot in Fig. 3a). The horizontal axis is shared between a and b. The vertical axis is shared between a and c, and labeled both in terms of $V_{IDT}$ and the charge amplitude $q_{SAW}$ induced on the SET island by the SAW. The colored traces in b and c are cross-sections of the two-dimensional data set, taken at the correspondingly colored lines in a. Measurement data in b and c are denoted by circles, and the calculated $\Gamma(q_g, q_{SAW})$ by solid curves. In b, the black circles show the SET transfer function in the absence of SAW, $\Gamma_0(q_{ind})=\Gamma(q_g,0)$, from which the theoretical function $\Gamma(q_g, q_{SAW})$ is computed. In the presence of a high-frequency SAW, the instantaneous induced charge varies between the extreme values of $q_{SAW}$, causing the SET response to smear. When the smearing extends over a full modulation period (blue arrow), the SET response to the static charge is suppressed (blue). As the SAW amplitude is increased further, the modulation curve changes polarity (red), and reaches maximal inverse amplitude at $q_{SAW} \approx 3/2\ e_{p-p}$ (red arrow). The pattern of polarity inversion repeats and continues, with a slow decay of the modulation amplitude as $q_{SAW}$ increases. This decay is seen more clearly in c, where $\Gamma(q_g, V_{IDT})$ is plotted for two fixed values of $q_g$. The DC component of $\Gamma$ increases slightly with $V_{IDT}$ (black dashes), due to rectified cross-talk. The theoretical curves have been adjusted by this experimental DC component. Apart from this, the only fitting parameter is the linear coefficient relating the applied voltage amplitude $V_{IDT}$ to the charge amplitude induced by the SAW on the SET island, $q_{SAW}$. This coefficient has the unit of capacitance, and we find that $C_{SAW}=q_{SAW}/V_{IDT}=81.5aF$ on resonance.

All measurements were done at temperatures below 200mK, where both the SET and the IDTs are superconducting, but the thermal energy is higher than the SAW phonon energy, $hf_{SAW} = 45mK \times k_B$. In our case, thermal phonons contribute to the noise background, but the sensitivity of the RF-SET is limited by amplifier noise, measured to $\delta q_g$=19µe$_{RMS}$/√Hz for charge induced through the gate. The sensitivity to $q_{SAW}$ was determined by using the SET as a heterodyne mixer[23], applying a local oscillator signal to $C_g$. This way, we obtained $\delta q_{SAW}$=25µe$_{RMS}$/√Hz. See Supplementary Methods I for details about the measurement scheme. For GaAs, the relation between vertical surface displacement and surface charge induced on the SET is

$$\frac{u_z}{q_{SAW}} = \frac{|c_z|}{|c_x|} \frac{\lambda}{2\pi e_{14} A_{SET}} = 1.2 \text{ am/µe}$$

where $c_x$ and $c_z$ are the ratios between surface displacement and electric SAW potential for the direction of SAW propagation and the surface normal, respectively (see Supplementary Methods III). Using this, we find an out-of-plane displacement sensitivity of $\delta u_z$=30am$_{RMS}$/√Hz. Better displacement sensitivities have been reported in other systems[24,25], but then by optical probing at considerably lower mechanical frequencies, where special cooling methods are required to reach the quantum regime. Previous reports on detection of weak SAW signals have used quantum dots[26] or scanning probes[27], and the improvement compared with these results is three orders of magnitude, due to the strong coupling of charge to the RF-SET and its high charge sensitivity.

The measured sensitivity can be compared with the ultimate limit set by thermal phonons and quantum fluctuations at the crystal surface, which we calculate to

$$\delta u_z = |c_z| \sqrt{\frac{2\pi h f}{y_0} \coth\left(\frac{hf}{2 k_B T}\right)}$$

where $y_0 = 3.1 \times 10^{-3}$ Ω$^{-1}$ is the characteristic SAW admittance of GaAs. This gives an ultimate displacement noise of $\delta u_{z,min}$= 0.046 am/√Hz set by quantum fluctuations in the crystal. The thermal phonons at 200mK increase this number by a factor of three to 0.14am/√Hz. See Supplementary Methods IV for details about this calculation.

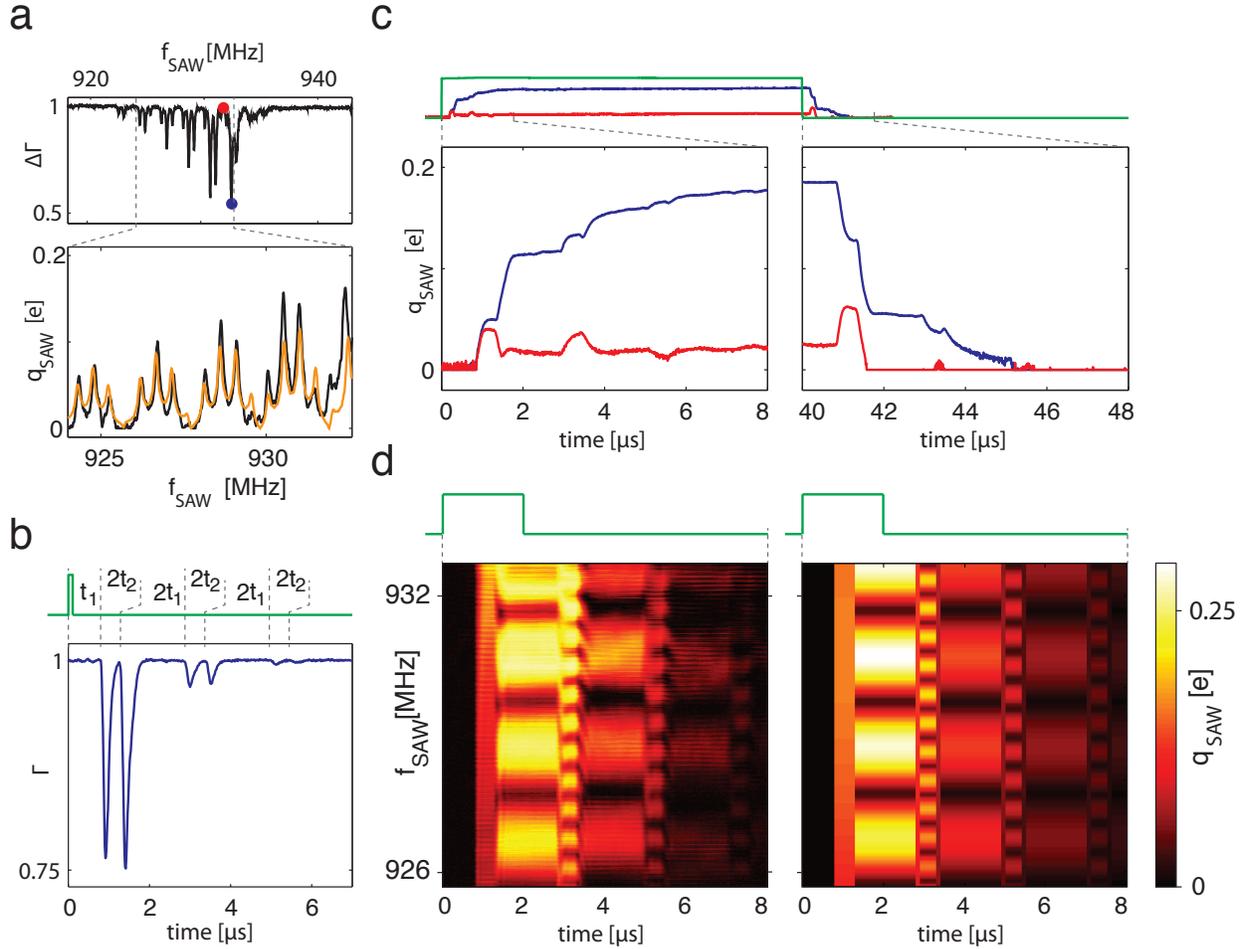

**Figure 3: Dynamic and frequency-dependent response of the SET to SAW**

a, Frequency response of the SAW cavity, as measured with the SET. Upper: Peak-to-peak amplitude of $\Gamma(q_g)$, denoted $\Delta\Gamma$, as a function of $f_{SAW}$. Lower: Partial zoom-up, converted to charge amplitude $q_{SAW}$ (black). The orange trace is a numerical simulation; see panel d and main text. Panels b, c and d show the averaged time-domain response of the SET, for SAW pulses with different properties. The envelope shapes of the applied pulses are shown above the plots in green, and the red and blue traces correspond to different SAW frequencies, shown with dots in panel a. The SET response follows the red curve in Fig. 2c. See also Supplementary Methods VII. b, Short microwave pulse (100ns): The acoustic wave is first detected at the SET after a delay of $t_1=d_1/v_{SAW}$, *i.e.* the SAW traversal time from IDT$_{left}$ to the SET. The next peak occurs when the SAW pulse has reflected at IDT$_{right}$ and returned back to the SET, *i.e.* after an additional $2t_2=2d_2/v_{SAW}$. In this reflection, no acoustic loss can be detected. The next peak, occurring after an additional $2t_1$, *i.e.* a round-trip from the SET to IDT$_{left}$, is significantly lower. c, SET response in charge units to long microwave pulses, with constructive (blue) and destructive (red) self-interference. In the blue and red traces above the main plots, the SET response is shown for the full duration of the pulses. The two main plots show zoom-ups of the beginning and end of the pulses. Here, the SET response forms plateaus, with lengths of $2t_1$ and $2t_2$, *i.e.* corresponding to round-trip times between the SET and the two IDTs. The first plateau (before the first SAW reflection) is similar for the two traces. However, after the first reflection, the left-going and right-going waves interfere at the SET. In the blue trace, this reflection and all successive ones add constructively. In the red trace, successive reflections add destructively, and the total amplitude remains low. d, Left: Measured SAW amplitude $q_{SAW}$ as a function of time and frequency, for 2μs long SAW pulses. Right: Theoretical model fitted to the measured data (see Supplementary Methods VI). The color axis is the same for the two plots, and the parameters derived from this fitting were used to calculate the steady-state frequency response shown in panel a (orange). Note: Where the conversion from SET response to induced charge $q_{SAW}$ results in a negative value due to noise, $q_{SAW}$ is truncated to zero.

Fig. 3a shows the strong dependence of the SET response on the SAW frequency, due to acoustic cavity resonances between the IDTs. The SAW frequency where the sensitivity measurement was done ($f_{SAW}$=932.407MHz) is indicated here by a blue dot.

With SET-to-IDT distances on the order of millimeters and a measurement bandwidth of 8MHz, we are not limited to steady-state measurements, but can study the SAW in the time domain as it traverses the chip. To illustrate this, we apply 100ns microwave pulses to $IDT_{left}$ and measure the response of the SET, as shown in Fig. 3b. Each pulse produces an isolated peak in the time trace each time it passes underneath the SET.

From the start of the SAW pulse, there is a delay before it is first detected by the SET, corresponding to the acoustic time of flight from $IDT_{left}$ to the SET. The lack of response in this time interval shows that there is no significant electrical crosstalk between the IDT and the SET. After an additional delay, corresponding to the round-trip time from the SET to the reflective $IDT_{right}$, the SAW pulse is detected again. The pulse continues back to $IDT_{left}$, reflects and returns to the SET, producing a third peak. The sequence of peaks continues as the pulse echoes back and forth, but the amplitude decreases significantly upon each reflection against $IDT_{left}$ (discussed below).

When the pulses are made longer, left-going and right-going SAW components, which emanate from the directly incident beam and successive reflections, interfere at the position of the SET. This is shown in Fig. 3c, where time traces were taken for two different SAW frequencies. The response of the SET changes each time it is reached by the SAW front, and stays constant during the next round-trip to an IDT. For constructive interference, we see the acoustic cavity ring up step by step after the microwave pulse is applied, and ring down after the pulse is removed. For destructive interference, successive reflections contribute with opposite signs, so that the total amplitude remains close to zero.

To verify the acoustic properties of the device, we measured the SET response as a function of time and SAW frequency for SAW pulses of 2μs duration (*i.e.* of a length around twice that of the cavity). This data set was then fitted with an acoustic model following Datta[1] (see Supplementary Methods VI). The agreement between measured data and simulation is good, as seen in Fig. 3d, and the fitted parameter values agree with tabulated values and independent measurements. The same parameters were used to compute the steady-state spectrum shown in Fig. 3a (orange). The simulation agrees reasonably well with the measurement also here, which supports the validity of the model.

The model confirms that all significant loss of SAW power occurs upon reflection against $IDT_{left}$, where the SAW loses ~80% of its power. We attribute this partly to back-transduction into the 50Ω transmission line, but scattering against local surface contaminants may also contribute.

In contrast with a phonon in a resonator, which has a single mode with a well-known energy, a propagating phonon is described as a wave packet with a distribution in both the time and frequency domains[28]. Such non-classical states have only recently been produced in the photonic case, and we cannot generate the phononic equivalents in our present setup. Instead, we use weak coherent pulses to demonstrate the detection sensitivity in our system, *i.e.* Poissonian superpositions of number states. Propagating phononic number states can in principle be generated by direct coupling to a superconducting qubit, as has been demonstrated for photons[29]. See further discussion below and in Supplementary Discussion II.

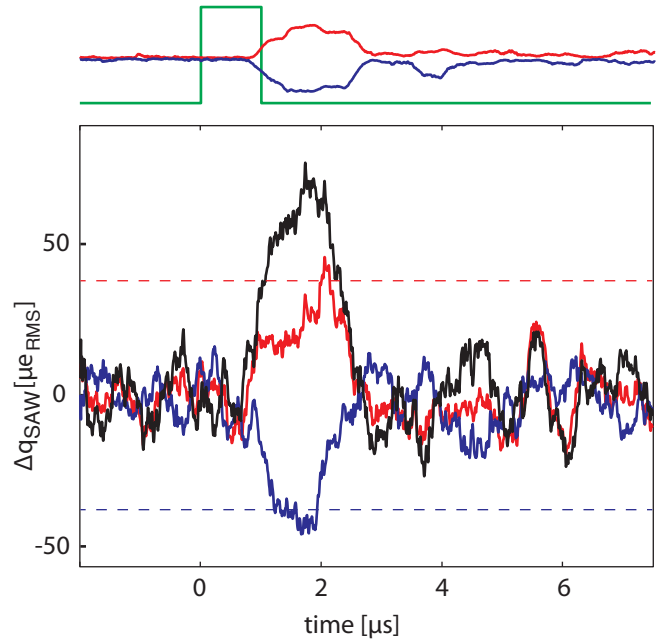

**Figure 4: Detecting SAW pulses on the single-phonon level.**

**Homodyne measurement of the SET response to weak SAW pulses at $f_{SAW}$ = 932.407MHz. Pulses of amplitude 1.5mV$_{RMS}$ were applied to the IDT through an attenuation chain of 80dB. The SAW power density is then $P_{SAW} = \beta_{gen} \times (150nV)^2/(50\Omega \times W_{IDT}) = 2.8 \times 10^{-19}$W/μm. With 1μs pulses and an SET width of 1.4μm, this corresponds to an average of ~0.6 phonons per pulse passing under the SET island. This sample has a wide IDT in order to avoid beam diffraction in the long acoustic cavity. Without this constraint, the IDT width is in principle arbitrary. The length of the pulse approximately equals that of the acoustic cavity, and the main hump in the data includes two traversals of the pulse past the SET. The red and blue traces both represent the change in RMS charge induced on the island by the SAW pulse, relative to the level before the pulse. Between the**

**two, the local oscillator phase was shifted by 180°, inverting the measured SET signal. Both traces were averaged $10^7$ times, and the black trace shows the difference between the two. The horizontal dashed lines show the expected pulse amplitudes of $\Delta q_{SAW}$=38μeRMS (see Supplementary Methods VII). Above the main panel, the green trace shows the envelope shape of the applied pulse. The red and blue traces show the same measurement as the main panel, but for much higher SAW amplitude.**

We are able to detect propagating acoustic pulses of extremely low amplitude, as shown in Fig. 4, where the SAW pulse was measured by homodyne mixing with a local oscillator signal applied to the SET gate[23]. For each pulse, the SAW energy passing under the SET island was $0.6h \times f_{SAW}$, i.e. corresponding to less than a single phonon on average. These pulses can clearly be detected with only two traversals in the acoustic cavity, by averaging over $10^7$ measurements.

We calculate the sensitivity enhancement required to reach the single-phonon level without averaging ("single shot") by several methods, and find that a factor $K_{SS} \approx 500$ is needed for a measurement like the one shown in Fig. 4. The calculations are given in Supplementary Methods IV.

The sensitivity of the SET can be improved in several different ways: The most straightforward is to increase the area of the SET island to pick up more of the surface charge. Optimization of the cold amplifier and tank circuit as well as optimized choices of geometry and substrate materials can give further improvements. Large additional gain can also be obtained from resonant enhancement of the signal, i.e. by embedding the probe in an acoustic cavity of much higher quality factor $Q$ than used in the present device. By the use of high-quality acoustic Bragg-mirrors, mechanical Fabry-Perot cavities can be constructed where $Q$ is dominated by the coupling to SAW modes outside the cavity. Such a cavity can be used as a capturing device for incoming phonons, embedding an SET or another probe or quantum circuit.

With increased coupling between the SET and the SAW, back-action from the SET also becomes an issue[30]. In our setup, the fundamental back-action arises from the SAW coupling piezoelectrically to the shot noise of quasiparticles tunnelling through the SET, and we cannot rule out an additional contribution due to SET self-heating.

We calculate that the SET emits and absorbs around $10^{-5}$ phonons per second and unit bandwidth due to shot noise, and this limits the quality factor of an acoustic cavity to $Q \ll 3 \times 10^5$ for the occupation to remain well below one phonon. To limit this effect, the SET should be driven on one of the quasiparticle/Cooper pair resonances[31,32], or be switched for a Quantum Capacitance Electrometer, which is similar to an SET in layout and coupling, but minimizes back-action since it is non-dissipative[33,34]. For stationary phonons in a closed cavity, such a quantum limited detector would enable reaching the optimal sensitivity discussed in Ref. [5]. For phonons propagating through a cavity, the measurement optimization is less well studied, and according to Helmer *et al.* the back-action gives an unavoidable reflection of incoming quanta, thus reducing the detection efficiency[35,36]. See Supplementary Methods V for a detailed discussion about back-action.

Although single-shot phonon detection is the ultimate prospect, interesting quantum mechanical experiments can be conducted with much lower measurement fidelity than this, as shown *e.g.* in the correlation measurements by Bozyigit *et al.*, where a signal-to-noise ratio substantially lower than unity sufficed to determine the quantum nature of a propagating photon field[37]. Studies of two-phonon interaction and of phononic crystals[38] in the quantum regime are other interesting experiments that can be done with an averaged read-out.

In the microwave regime, a superconducting qubit is an ideal generator of non-classical photonic states in electrical cavities and transmission lines. It is interesting to consider whether such states can be produced also in a SAW device, and their quantum nature confirmed using on-chip probes or linear microwave amplifiers. One can easily imagine an experiment similar to that of Ref. [6], where the "quantum drum" is replaced by a one-port or two-port SAW resonator, coupled as an impedance element to a superconducting qubit through an electrical transmission line. This has possible advantages due to the high mechanical performance of SAW devices at low temperature.

However, we can also consider a qubit that couples directly to the mechanical wave, in the same way as the RF-SET does in the presented experiment. Qubits of the "transmon" type[39] are well suited for this, due to their geometry and way of interacting with their environment. A transmon consists of a Josephson junction with Josephson energy $E_J$, shunted by a large geometric capacitance, which dominates the total capacitance $C_Q$ of the transmon. In circuit QED, the transmon is normally positioned in the dielectric gap of a coplanar cavity or transmission line, the electric field of which interacts with the polarization of the qubit. If the geometric capacitance is instead fashioned into the shape of an IDT, the transmon polarization couples directly to SAW propagating on the underlying substrate, in a similar way to the SET. By comparing two equivalent circuits, one acoustic and one all-electric, we find that the acoustic coupling can be as strong as the electric one, for $f_0$=5GHz and $E_J/E_C = 40$, where $E_C \approx e^2/(2C_Q)$, along with other reasonable sample parameters (See Supplementary Discussion II). The strong coupling suggests that phononic versions are feasible of experiments that have been demonstrated in the field of circuit QED[40]. These include the generation of single phonons by example of Houck *et al.*[29], as well as experiments where single quanta interact with artificial atoms in open transmission lines[41,42].

**Acknowledgements:** We thank Konrad W. Lehnert and Jari Kinaret for discussions, and Tord Claeson for commenting on the manuscript. Financial support by the EU FW6 grant "ACDET II", the European Research Council, the Swedish VR and the Wallenberg foundation is gratefully acknowledged.